\documentclass[11pt,a4paper]{article}
\usepackage{jcappub}
\usepackage{euscript,epsfig,amsmath,amssymb}
\usepackage{amsfonts,latexsym}

\newcommand{\be}[1]{\begin{equation}\label{#1}}
\newcommand{\ee}{\end{equation}}
\newcommand{\ba}[1]{\begin{eqnarray}\label{#1}}
\newcommand{\ea}{\end{eqnarray}}
\newcommand{\rf}[1]{(\ref{#1})}
\newcommand{\nn}{\nonumber}
\renewcommand{\theequation}{\arabic{section}.\arabic{equation}}

\newcommand{\const}{\mbox{\rm const}\,}

\begin{document}

\title{Are dark energy models with variable EoS parameter $w$ compatible with\\ the late inhomogeneous Universe?}

\author{\"{O}zg\"ur Akarsu$^{1}$,}
\author{Mariam Bouhmadi-L\'opez$^{2,3,4,5}$,}
\author{Maxim Brilenkov$^{6}$,}
\author{Ruslan Brilenkov$^{6}$,}
\author{Maxim Eingorn$^{7}$,}
\author{and Alexander Zhuk$^{8}$}

\affiliation{$^{1}$Department of Physics, Ko\c{c} University, 34450 Sar{\i}yer, {\.I}stanbul, Turkey\\}

\affiliation{$^{2}$Departamento de F\'{i}sica, Universidade da Beira Interior, 6200 Covilh\~a, Portugal \\}

\affiliation{$^{3}$Centro de Matem\'atica e Aplica\c{c}\~oes da Universidade da Beira Interior (CMA-UBI),\\ 6200 Covilh\~{a}, Portugal\\}

\affiliation{$^{4}$Department of Theoretical Physics, University of the Basque Country
UPV/EHU,\\ P.O. Box 644, 48080 Bilbao, Spain\\}

\affiliation{$^{5}$IKERBASQUE, Basque Foundation for Science, 48011 Bilbao, Spain\\}

\affiliation{$^{6}$Department of Theoretical Physics, Odessa National University,\\ Street Dvoryanskaya 2, Odessa 65082, Ukraine\\}

\affiliation{$^{7}$North Carolina Central University, CREST and NASA Research Centers,\\ Fayetteville st. 1801, Durham, North Carolina 27707, U.S.A.\\}

\affiliation{$^{8}$Astronomical Observatory, Odessa National University,\\ Street Dvoryanskaya 2, Odessa 65082, Ukraine\\}

\emailAdd{oakarsu@ku.edu.tr} \emailAdd{mbl@ubi.pt On leave of absence from UPV and IKERBASQUE}
\emailAdd{maxim.brilenkov@gmail.com}
\emailAdd{ruslan.brilenkov@gmail.com}
\emailAdd{maxim.eingorn@gmail.com} \emailAdd{ai.zhuk2@gmail.com}

\abstract{We study the late-time evolution of the Universe where dark energy (DE) is presented by a barotropic fluid on top of cold dark matter (CDM). We also
take into account the radiation content of the Universe. Here by the late stage of the evolution we refer to the epoch where CDM is already clustered into
inhomogeneously distributed discrete structures (galaxies, groups and clusters of galaxies). Under this condition the mechanical approach is an adequate tool to
study the Universe deep inside the cell of uniformity. More precisely, we study scalar perturbations of the FLRW metric due to inhomogeneities of CDM as well as
fluctuations of radiation and DE. For an arbitrary equation of state for DE we obtain a system of equations for the scalar perturbations within the mechanical
approach. First, in the case of a constant DE equation of state parameter $w$, we demonstrate  that our method singles out the cosmological constant as the only
viable dark energy candidate. Then, we apply our approach to variable equation of state parameters in the form of three different linear parametrizations of $w$,
e.g., the Chevallier-Polarski-Linder perfect fluid model. We conclude that all these models are incompatible with the theory of scalar perturbations in the late
Universe.}

\maketitle

\flushbottom

\section{Introduction}

Explaining the accelerated expansion of the late Universe is one of the greatest challenges in modern cosmology. The $\Lambda$CDM model is successfully  described
by the Planck data \cite{Planck}. However, the nature of the cosmological constant is still unclear. In fact, there are about 120 orders of magnitude between the
observed value of the cosmological constant and the theoretically expected value. The latter is related to the vacuum energy density. The cancellation mechanism
that would reduce this large theoretical value of the cosmological constant is still a mystery \cite{Dolgov}. In addition, the $\Lambda$CDM model (as well as a
lot of other dark energy models) faces the coincidence problem, i.e. why is the cosmological constant at present of the same order of magnitude as the dark matter
energy density? To solve these problems, different dynamical dark energy models were proposed.
For example, such dark energy
can be modeled by a barotropic perfect fluid (i.e. a fluid whose density is a function of its pressure only) with a proper equation of state (EoS). The
Chevallier-Polarski-Linder (CPL) parametrization of the EoS \cite{ChevPol,Linder} is often used in literature (see, e.g., the recent papers
\cite{Lazkoz,CPL1,CPL2}). In this approach the parameter of the EoS is a simple linear function with respect to the scale factor $a$ of the Universe: ${w}(a) =
{w}_0+{w}_1(1-a/a_0)$, where ${w}_{0,1}$ are free parameters of the model and $a_0$ is the present day value of the scale factor. The energy density
$\varepsilon_{\mathrm{CPL}}$ of this fluid behaves as $\varepsilon_{\mathrm{CPL}} \sim a^{-{3(1+{w}_0+{w}_1)}}$ at early times when $a\ll a_0$, and when $a$ is
close to $a_0$ as well. Therefore, if ${w}_{0,1}$ are close to $-1$ and $0$ respectively, then this barotropic fluid mimics the cosmological constant.
Recently BOSS collaboration gave the constraints on the parameters of this EoS from the Planck+BAO+SN data as $w_0=-0.93\pm 0.11$ $(1\sigma)$ and
$w_1=-0.2\pm0.4$ $(1\sigma)$ \cite{BOSS14}. In~\cite{OzAl} an alternative form of the EoS was suggested following from the linear time parametrization
$w(t)=w_{0}+w_{1}\left(1-t/t_{0}\right)$. Using the same data and method with \cite{BOSS14}, the authors found that the Planck+BAO+SN data predict for
this model ${w}_0 =-0.99\pm 0.06$ $(1\sigma)$ and $-1<{w}_0+{w}_1 <-0.42$ $(2\sigma)$.

It is of interest to investigate these models from the point of their compatibility with other cosmological and astrophysical data, as well as with the
theory of cosmological perturbations, in particular, with the theory of scalar perturbations in the late Universe. At the late stage of the Universe
evolution when inhomogeneities (such as galaxies and their groups) are already formed, the hydrodynamic approach is inadequate. Here the mechanical
approach \cite{EZcosm1,EKZ2} is more appropriate. It works well inside the cell of uniformity \cite{EZcosm2} and provides us a good tool to investigate
scalar perturbations for different cosmological models (see, e.g., \cite{BUZ1,Laslo2,Novak}). Therefore, it makes sense to study cosmological models
filled with perfect fluids which have linear EoS parametrizations and investigate the compatibility of these models with the mechanical approach.
Obviously, the above-mentioned forms of the EoS are among the simplest ones. In our paper, for completeness, we also consider a linear parametrization
with respect to the redshift \cite{Cooray}: $w(z)=w_{0}-w_{1}z$. As a result, we show that all three considered models are incompatible with the theory of
scalar perturbations in the late Universe. This means that such parametrizations can be used only for a limited period of time, starting, e.g., from last
scattering till present. That is they follow from some more general EoS and for late times (in the future) they should be replaced by other expressions.

The paper is structured as follows. In Sec. 2 we derive the basic equations for scalar perturbations in the framework of the mechanical approach in the
case of a barotropic perfect fluid with an arbitrary EoS $p_{\mathrm{X}}= p_{\mathrm{X}}( \varepsilon_{\mathrm{X}})$. Then, in Secs. 3, 4 and 5 we
demonstrate that the enumerated DE EoS parametrizations linear in $a$, $z$ and $t$, respectively, are incompatible with the scalar perturbations theory.
The main results are briefly summarized in concluding Sec. 6. In Appendix we consider a particular case of a barotropic perfect fluid with a constant EoS
parameter. Here we investigate the conservation equations for scalar perturbations in the mechanical approach and demonstrate its self-consistency.

\section{Basic equations}

\subsection{Background}

We consider the Universe at late stages of its evolution when galaxies and clusters of galaxies have already formed. At scales of the order of approximately 190
Mpc \cite{EZcosm2}, being much larger than the characteristic distance between these inhomogeneities, the Universe is well described by the homogeneous and
isotropic FLRW metric and the Friedmann equations

\be{2.1}
\frac{3\left(\mathcal{H}^2+\mathcal{K}\right)}{a^2}= \kappa\left(\overline{T}^0_{0\mathrm{\, (CDM)}}+\overline \varepsilon_{\mathrm{rad}} + \overline
\varepsilon_{\mathrm{X}} \right) +\Lambda
\ee
and
\be{2.2}
\frac{2\mathcal{H}'+\mathcal{H}^2+\mathcal{K}}{a^2}=-\kappa\left(\overline p_{\mathrm{rad}}+ \overline p_{\mathrm{X}}\right) + \Lambda\, ,
\ee
where ${\mathcal H}\equiv a'/a\equiv (da/d\eta)/a$, \ $\kappa\equiv 8\pi G_N/c^4$ ($c$ is the speed of light and $G_N$ is the Newtonian gravitational constant)
and $\mathcal K=-1,0,+1$ stands for open, flat and closed Universes, respectively. Conformal time $\eta$ and synchronous time $t$ are connected as $cdt=a d\eta$.
Further, $\overline T^{i}_{k\mathrm{\, (CDM)}}$ is the average energy-momentum tensor of the pressureless (dust-like) matter. For such matter the energy density
$\overline T^{0}_{0\mathrm{\, (CDM)}} =\overline \rho_{\mathrm{\, c}} c^2/a^3$ is the only nonzero component, here $\overline \rho_{\mathrm{\, c}}=\mbox{const}$
is the average comoving rest mass density \cite{EZcosm1}. As usual, for radiation we have the EoS $\overline p_{\mathrm{rad}}=(1/3)\overline
\varepsilon_{\mathrm{rad}}$. The Universe is also supposed to be filled with a barotropic perfect fluid with the EoS $\overline p_{\mathrm{X}}=\overline
p_{\mathrm{X}}(\overline \varepsilon_{\mathrm{X}})$ corresponding to DE. For generality, we also included the cosmological constant $\Lambda$ into the above
equations. Obviously, if the barotropic fluid is the reason of the late time acceleration of the Universe, then the cosmological constant is not necessary. It can
be easily seen that the presence (or absence) of the cosmological constant does not affect the scalar perturbation analysis. For example, one can write down the
Einstein equations with an upper index and a lower one. Then, the $\Lambda$ term will correspond to the contribution $\Lambda\delta^i_k$ whose perturbations are
zero at any order{\footnote{The conservation equation clearly shows that the $\Lambda$ term can vary only in the case of interaction with other matter (see, e.g.,
\cite{Kiefer}). However, this is out of the scope of our model.}}. However, the dynamical behavior of the Universe (i.e. the form of the scale factor $a$) depends
on $\Lambda$.

From Eqs. \rf{2.1} and \rf{2.2} we can easily get the following auxiliary equation:
\be{2.3}
\frac{2}{a^2}\left(\mathcal{H}'-\mathcal{H}^2-\mathcal{K}\right)= -\kappa\left(\overline{T}^0_{0\mathrm{\, (CDM)}}+\overline
\varepsilon_{\mathrm{rad}} +\overline \varepsilon_{\mathrm{X}} + \overline p_{\mathrm{rad}}+ \overline p_{\mathrm{X}}\right)\, .
\ee

\

\subsection{Scalar perturbations}

Deep inside the cell of uniformity the Universe is highly inhomogeneous. Here the mechanical approach is more adequate than the hydrodynamical one
\cite{EZcosm1,EZcosm2}. In the framework of the mechanical approach galaxies, dwarf galaxies and clusters of galaxies (composed of baryonic and dark
matter) are considered as separate compact objects. Moreover, at distances much greater than their characteristic sizes they can be described well as
point-like matter sources. This is a generalization of the well-known astrophysical approach \cite{Landau} (see \S 106) for the case of the dynamical
cosmological background. Usually, the gravitational fields of galaxies are weak and their peculiar velocities are much smaller than the speed of light.
These inhomogeneities together with fluctuations of other matter sources result in scalar perturbations of the FLRW metric \cite{Mukhanov,Rubakov}. In the
mechanical approach and
in the conformal Newtonian gauge, the gravitational potential (i.e. the perturbation of the metric coefficient $g_{00}$) $\Phi \sim O(1/c^2)$ satisfies the
following system of equations (see \cite{EZcosm1,EZcosm2,Laslo2} for details):
\ba{2.4} &{}&\Delta\Phi-3\mathcal{H}(\Phi'+\mathcal{H}\Phi)+3\mathcal{K}\Phi=\frac{1}{2}\kappa a^2\left(\delta T^0_{0\mathrm{\, (CDM)}}+
\delta\varepsilon_{\mathrm{X}}+
\delta\varepsilon_{\mathrm{rad1}}+\delta\varepsilon_{\mathrm{rad2}}\right)\, ,\\
\label{2.5}
&{}&\frac{\partial}{\partial x^\beta}(\Phi'+\mathcal{H}\Phi)=0\, ,\\
\label{2.6} &{}&\Phi''+3\mathcal{H}\Phi'+(2\mathcal{H}'+\mathcal{H}^2)\Phi-\mathcal{K}\Phi=\frac{1}{2}\kappa a^2\left(\delta p_{\mathrm{X}}+ \delta
p_{\mathrm{rad1}}+\delta p_{\mathrm{rad2}}\right)\, , \ea
where the Laplace operator $\triangle$ is defined with respect to the conformal spatial metric, and $x^{\beta},\, \beta=1,2,3$, are the comoving spatial
coordinates. From Eq. \rf{2.5} we get immediately that
\be{2.7}
\Phi(\eta,{\bf r})=\frac{\varphi({\bf r})}{c^2a(\eta)}\, ,
\ee
where $\varphi({\bf r})$ is a function of all comoving spatial coordinates, and we have introduced $c^2$ in the denominator for convenience. In the surrounding of
an inhomogeneity, the comoving potential $\varphi({\bf r})\sim 1/r$ \cite{EZcosm1,EZcosm2,BUZ1}, and the nonrelativistic gravitational potential $\Phi(\eta,{\bf
r})\sim 1/(a r)=1/R$, where $R=ar$ is the physical distance. Hence, $\Phi$ has the correct Newtonian limit near the inhomogeneities.

Concerning the fluctuations of the matter sources, $\delta T^0_{0\mathrm{\, (CDM)}}$ is related to the fluctuation of the energy density of dust-like matter and
has the form{\footnote{It can be easily seen from Eq. (2.23) in \cite{EZcosm1} that these fluctuations read $\delta T^0_{0\mathrm{\,
(CDM)}}=\delta\rho_{\mathrm{c}}\, c^2/a^3 + (\overline{\rho}_{\mathrm{c}}c^2/a^3)[3\Phi +\alpha\Phi^2+\beta\Phi^3+...] $ where $\alpha$ and $\beta$ are some
constants. Taking into account Eq. \rf{2.7}, we get $\Phi/a^3 \sim O(1/a^4)$, $\Phi^2/a^3 \sim O(1/a^5)$, $\Phi^3/a^3 \sim O(1/a^6)$ and so on. Since we consider
linear perturbations, we should drop the terms $\Phi^2, \Phi^3, \ldots\; $. Therefore, in what follows, we will keep terms up to the order $O(1/a^4)$ inclusive.}}
\cite{EZcosm1}
\be{2.8}
\delta T^0_{0\mathrm{\, (CDM)}}=\frac{\delta\rho_{\mathrm{c}}\, c^2}{a^3}+\frac{3\overline{\rho}_{\mathrm{c}}\, c^2\Phi}{a^3}\, ,
\ee
where $\delta\rho_{\mathrm{\, c}}$ is the difference between the real and average comoving rest mass densities: $\delta\rho_{\mathrm{ c}} = \rho_{\mathrm{\,
c}}-\overline\rho_{\mathrm{\, c}}$. We also split the fluctuations of radiation into two parts. The part labeled by ``rad1'' is caused by the inhomogeneities of
dust-like matter (e.g., by galaxies and their groups), and the part labeled by ``rad2'' is related to fluctuations of the additional perfect fluid (labeled by
$X$). Of course, in the experiments we measure the total fluctuation of radiation. However, inhomogeneities associated with different matter sources can
contribute to this total fluctuation separately. With respect to the additional perfect fluid, as we will see below, such contributions may take place directly
via nonzero $\delta\varepsilon_{\mathrm{rad2}}$ (cf. Eq.~\rf{2.22}). This happens if the inhomogeneities of the additional perfect fluid are distributed
independently of the fluctuations of CDM \cite{Laslo2}. However, there is also the possibility that the fluctuations of this perfect fluid nest on the
inhomogeneities of CDM (such as galaxies, groups of galaxies, etc) screening the latter. In such models, $\delta\varepsilon_{\mathrm{rad2}}=0$,
$\delta\varepsilon_{\mathrm{X}}\sim \varphi$ and Eq. \rf{2.17} (see below) is reduced (for physically consistent cases) to the Helmholtz  equation \cite{BUZ1}.
Then, the fluctuations of the additional perfect fluid contribute to the fluctuations of radiation indirectly via the comoving gravitational potential $\varphi
(\bf r)$ (cf. Eq.~\rf{2.10}) which is a solution of the corresponding Helmholtz equation.  Given these two possible scenarios, we include
$\delta\varepsilon_{\mathrm{rad2}}$ in our equations. Only a detailed analysis of a given model reveals which of these two scenarios is realized.

For both of these contributions
we have the same equations of state:
\be{2.9}
\delta p_{\mathrm{rad1}}=(1/3)\delta\varepsilon_{\mathrm{rad1}},\quad \delta
p_{\mathrm{rad2}}=(1/3)\delta\varepsilon_{\mathrm{rad2}}\, .
\ee
We have shown in~\cite{EZcosm2} that $\delta\varepsilon_{\mathrm{rad1}}$ has the form
\be{2.10}
\delta\varepsilon_{\mathrm{rad1}}=-\frac{3\overline{\rho}_{\mathrm{c}}\,\varphi}{a^4}\, .
\ee
It is worth noting that $\delta\varepsilon_{\mathrm{rad2}}$ should also behave as $1/a^4$ (see Appendix).

Let us now analyze Eqs. \rf{2.4} and \rf{2.6}. Taking into account Eq.~\rf{2.7}, we can rewrite them as follows:
\ba{2.11} &{}&\Delta\Phi+3\mathcal{K}\Phi=
\frac{1}{2}\kappa a^2\left(\delta T^0_{0\mathrm{\, (CDM)}}+\delta\varepsilon_{\mathrm{X}}+\delta\varepsilon_{\mathrm{rad1}}+
\delta\varepsilon_{\mathrm{rad2}}\right)\, ,\\
\label{2.12} &{}&\left(\mathcal{H}'-\mathcal{H}^2-\mathcal{K}\right)\Phi= \frac{1}{2}\kappa a^2\left(\delta p_{\mathrm{X}}+\delta p_{\mathrm{rad1}}+\delta
p_{\mathrm{rad2}}\right)\, . \ea
Substituting Eqs.~\rf{2.8} and \rf{2.10} into the right hand side (rhs) of Eq.~\rf{2.11}, we obtain
\be{2.13} \Delta\Phi+3\mathcal{K}\Phi= \frac{1}{2}\kappa a^2\left(\frac{\delta\rho_{\mathrm{c}}\,
c^2}{a^3}+\delta\varepsilon_{\mathrm{X}}+\delta\varepsilon_{\mathrm{rad2}}\right)\, . \ee
On the other hand, substituting \rf{2.3} into the left hand side (lhs) of Eq.~\rf{2.12}, we get the equation
\be{2.14}
-\left(\frac{\overline \rho_{\mathrm{ c}}\, c^2}{a^3}+\overline \varepsilon_{\mathrm{rad}} +\overline
\varepsilon_{\mathrm{X}} + \overline p_{\mathrm{rad}}+ \overline p_{\mathrm{X}}\right)\Phi
= \delta p_{\mathrm{X}}+\delta p_{\mathrm{rad1}}+\delta p_{\mathrm{rad2}}\, .
\ee
Now, for consistency, we must take into account terms up to the order $O(1/a^4)$ inclusive. This is the accuracy of our investigations. Then, the terms
$\overline \varepsilon_{\mathrm{rad}} \Phi$ and $\overline p_{\mathrm{rad}} \Phi$ of the order $O(1/a^5)$ (cf. Eq.~\rf{2.7}) should be dropped:
\be{2.15}
-\left(\frac{\overline \rho_{\mathrm{ c}}\, c^2}{a^3} +\overline
\varepsilon_{\mathrm{X}} +  \overline p_{\mathrm{X}}\right)\Phi
= \delta p_{\mathrm{X}}+\delta p_{\mathrm{rad1}}+\delta p_{\mathrm{rad2}}\, .
\ee
This equation can be rewritten in the form
\be{2.16}
-\left(\overline
\varepsilon_{\mathrm{X}} +  \overline p_{\mathrm{X}}\right)\Phi
= \delta p_{\mathrm{X}}+\frac{1}{3}\delta \varepsilon_{\mathrm{rad2}}\, ,
\ee
where we have used Eqs. \rf{2.9} and \rf{2.10}.

Therefore, Eqs. \rf{2.4} and \rf{2.6} are reduced to the following system:
\ba{2.17}
&{}&\triangle\varphi+3\mathcal{K}\varphi=\frac{\kappa c^4}{2}\delta\rho_{\mathrm{ c}}+\frac{\kappa c^2 a^3}{2}\delta\varepsilon_{\mathrm{X}}+\frac{\kappa
c^2a^3}{2}\delta\varepsilon_{\mathrm{rad2}}\, ,\\
\label{2.18} &{}&-\left(\overline{\varepsilon}_{\mathrm{X}}+\overline{p}_{\mathrm{X}}\right)\frac{\varphi}{c^2a}=\delta
p_{\mathrm{X}}+\frac{1}{3}\delta\varepsilon_{\mathrm{rad2}}\, . \ea

Obviously, DE described by the barotropic perfect fluid must satisfy this system at any moment of the cosmic time in the late Universe. Such analysis
should be performed for each EoS $\overline p_{\mathrm{X}}=\overline p_{\mathrm{X}}(\overline \varepsilon_{\mathrm{X}})$ individually.

For example, we can consider a perfect fluid with a constant parameter in the linear EoS:
\be{2.19}
\bar p_{\mathrm{X}} = {w} \bar \varepsilon_{\mathrm{X}}\, , \quad {w} \neq 0\, .
\ee
We exclude the case ${w} =0$ since the CDM component was already taken into account in Eq.~\rf{2.1}{\footnote{In addition, if $w=0$: (i) we get from Eq. \rf{2.18}
that $\delta\varepsilon_{\mathrm{rad2}}=-\overline{\varepsilon}_{\mathrm{X}}[3\varphi/(c^2a)]=-3\overline A\varphi/(c^2a^4)$ which is similar to Eq. \rf{2.10},
and (ii) Eq. \rf{2.17} implies $\delta\varepsilon_{\mathrm{X}}=B/a^3 -\delta\varepsilon_{\mathrm{rad2}}$ in full analogy with Eq. \rf{2.8} (for $B \equiv
\delta\rho_{\mathrm{ cX}}c^2$). Obviously, we can combine both of these dust components into a single one. Therefore, we would be back to the model with CDM and a
$\Lambda$ term (in the background equations \rf{2.1} and \rf{2.2}).}}. From the conservation equation we have
\be{2.20}
\bar \varepsilon_{\mathrm{X}} = \frac{\bar A}{a^{3(1+{w})}}\, .
\ee
Further, for fluctuations of this matter we have
\be{2.21}
\delta p_{\mathrm{X}} = {w} \delta\varepsilon_{\mathrm{X}}\, .
\ee
Therefore, Eq. \rf{2.18} gives
\be{2.22}
\delta\varepsilon_{\mathrm{X}} = -\frac{\varphi}{c^2a}\frac{1+{w}}{{w}}\bar\varepsilon_{\mathrm{X}} - \frac{1}{3{w}}
\delta\varepsilon_{\mathrm{rad2}}\, .
\ee
Substituting this relation into \rf{2.17}, we arrive at the following equation:
\be{2.23}
\triangle\varphi+3\mathcal{K}\varphi-\frac{\kappa c^4}{2}\delta\rho_{\mathrm{ c}}=
-\frac{\kappa \varphi}{2}\frac{1+{w}}{{w}}a^2\bar\varepsilon_{\mathrm{X}}
+\frac{\kappa
c^2a^3}{2}\left(1-\frac{1}{3{w}}\right)\delta\varepsilon_{\mathrm{rad2}}\, .
\ee
The lhs of this equation does not depend on time{\footnote{We would like to remind that the quantities $\varphi$ and $\delta\rho_{\mathrm{ c}}$ are the comoving
ones \cite{EZcosm1}. Therefore, within the adopted accuracy when both nonrelativistic and weak field limits are applied, they do not depend explicitly on time
\cite{EZcosm2}.}}. Therefore, the same should hold true for its rhs. Because $\delta\varepsilon_{\mathrm{rad2}} \sim 1/a^4$, we must demand that
$\delta\varepsilon_{\mathrm{rad2}}\equiv 0$ (there is no possibility that the term with $\bar\varepsilon_{\mathrm{X}}$ and the term with
$\delta\varepsilon_{\mathrm{rad2}}$ cancel each other). Therefore, there are only two values of ${w}$ which do not contradict this equation. They are ${w} =-1$
and ${w} = -1/3$. The former case reduces to the standard $\Lambda$CDM model considered in \cite{EZcosm1}. The latter case ${w}=-1/3$ was considered in detail in
\cite{BUZ1} in the absence of radiation. The same conclusions with respect to the model with constant ${w}$ can be obtained from the conservation equations (see
Appendix). Obviously, the dark energy component is smooth and homogeneous in the background equations where matter/energy is considered in the averaged form.
However, it does not have to be smooth and homogeneous a priori. The conclusion of the presence or absence of dark energy fluctuations follows from the analysis
of Einstein equations for perturbations. For example, for the $\Lambda$CDM model (i.e. $w=-1$) fluctuations are absent, as it is easily seen from Eq. \rf{2.22}
where $\delta\varepsilon_{\mathrm{rad2}}=0$ (see also \cite{EZcosm1}), so dark energy represented by the cosmological constant $\Lambda$ is truly homogeneous,
while for the model with $w=-1/3$, fluctuations are nonzero \cite{BUZ1}, so there are inhomogeneities in the dark energy distribution over the Universe. It is
clear that in the case of the inhomogeneous distribution, the average value of fluctuations should be equal to zero: $\overline{\delta\varepsilon_{\mathrm{X}}}=0$
(see, e.g., \cite{BUZ1} where it is shown explicitly).

We next investigate the models with dynamical ${w}$.


\section{CPL model}

First, we consider the CPL model (i.e. $\mathrm{X} \equiv \mathrm{CPL}$) and the corresponding EoS \cite{ChevPol,Linder}:
\be{3.1}
\overline p_{\mathrm{CPL}}={w}(a)\overline\varepsilon_{\mathrm{CPL}}\, , \quad
{w}(a)={w}_{0}+{w}_{1}\left(1-\frac{a}{a_{0}}\right)\, ,
\ee
where $a_0$ represents the scale factor of the Universe at the present time. We consider this EoS to be valid for a definite period of time in the past
(e.g., from last scattering \cite{Linder}). We also suppose that this EoS is still valid for some period of time in the future.

From the conservation equation in the form
\be{3.2} d(\overline\varepsilon_{\mathrm{CPL}} a^{3})+\overline p_{\mathrm{CPL}}d(a^{3})=0 \ee
we easily get the well known formula
\be{3.3}
\overline\varepsilon_{\mathrm{CPL}}=Aa^{-3\left(1+{w}_{0}+{w}_{1}\right)} e^{3{w}_{1}a/a_{0}} \, ,
\ee
where $A$ is a constant of integration. Therefore,
\be{3.4}
\overline p_{\mathrm{CPL}}+\overline\varepsilon_{\mathrm{CPL}} =
Aa^{-3\left(1+{w}_{0}+{w}_{1}\right)} e^{3{w}_{1}a/a_{0}}\left[1+{w}_{0}+{w}_{1}\left(1-\frac{a}{a_{0}}\right)\right]\, .
\ee
Additionally, with the help of Eq. \rf{3.3}, we obtain the expression for the fluctuation of the CPL fluid pressure:
\ba{3.5}
\delta p_{\mathrm{CPL}} &=& \frac{\partial \left({w}\overline\varepsilon_{\mathrm{CPL}}\right)}{\partial \overline\varepsilon_{\mathrm{CPL}}}
\delta\varepsilon_{\mathrm{CPL}}
=\left({w}-\frac{{w}_1\overline\varepsilon_{\mathrm{CPL}}}{a_0}\frac{da}{d\overline\varepsilon_{\mathrm{CPL}}}\right)
\delta\varepsilon_{\mathrm{CPL}} \nn \\
&=& \left[{w}_{0}+{w}_{1}\left(1-\frac{a}{a_{0}}\right)-
\frac{1}{3}\frac{{w}_1}{{w}_1-\frac{a_0}{a}\left(1+{w}_{0}+{w}_{1}\right)}\right]\delta\varepsilon_{\mathrm{CPL}}\, .
\ea
Therefore, Eq. \rf{2.18} reads
\ba{3.6}
-\frac{A\varphi}{c^2}a^{-3\left(1+{w}_{0}+{w}_{1}\right)-1} e^{3{w}_{1}a/a_{0}}\left[1+{w}_{0}+{w}_{1}\left(1-\frac{a}{a_{0}}\right)\right]&{}&
\nn \\
=\left[{w}_{0}+{w}_{1}\left(1-\frac{a}{a_{0}}\right)-
\frac{1}{3}\frac{{w}_1}{{w}_1-\frac{a_0}{a}\left(1+{w}_{0}+{w}_{1}\right)}\right]\delta\varepsilon_{\mathrm{CPL}}+
\frac{1}{3}\delta\varepsilon_{\mathrm{rad2}}\,
,&{}& \ea
and for $\delta\varepsilon_{\mathrm{CPL}}$ we have
\ba{3.7}
\delta\varepsilon_{\mathrm{CPL}} &=&
-\frac{A\varphi}{c^2}a^{-3\left(1+{w}_{0}+{w}_{1}\right)-1}
e^{3{w}_{1}\beta}\frac{\left(1+{w}_{0}+{w}_1\left(1-\beta\right)\right)^2}{{w}_{0}+{w}_1\left(1-2\beta/3\right)+\left({w}_{0}+
{w}_1\left(1-\beta\right)\right)^2}\nn \\
&-& \frac{1}{3}\delta\varepsilon_{\mathrm{rad2}}\frac{1+{w}_{0}+{w}_1\left(1-\beta\right)}{{w}_{0}+{w}_1\left(1-2\beta/3\right)+\left({w}_{0}+
{w}_1\left(1-\beta\right)\right)^2}\, ,
\ea
where for convenience we have introduced the notation $\beta \equiv a/a_0$. Consequently, Eq.~\rf{2.17} takes the form
\ba{3.8}
&{}&\triangle\varphi+3\mathcal{K}\varphi=\frac{\kappa c^4}{2}\delta\rho_{\mathrm{ c}}-\frac{A \kappa \varphi}{2}a^{-3\left(1+{w}_{0}+{w}_{1}\right)+2}
e^{3{w}_{1}\beta}\frac{\left(1+{w}_{0}+{w}_1\left(1-\beta\right)\right)^2}{{w}_{0}+{w}_1\left(1-2\beta/3\right)+\left({w}_{0}+
{w}_1\left(1-\beta\right)\right)^2}\nn \\
&+&\frac{\kappa c^2a^3}{2}\delta\varepsilon_{\mathrm{rad2}}\left[
1-\frac{1}{3}\frac{1+{w}_{0}+{w}_1\left(1-\beta\right)}{{w}_{0}+{w}_1\left(1-2\beta/3\right)+\left({w}_{0}+
{w}_1\left(1-\beta\right)\right)^2}\right]\, .
\ea

The particular case ${w}_1=0$, i.e.  ${w} = {w}_0 =\mathrm{const}$, was considered  at the very end of the previous section. Now, we turn to the general case
${w}_1\neq 0$ and consider Eq. \rf{3.8}. Here the lhs and the term containing $\delta\rho_{\mathrm{c}}$ are independent of time. Therefore, either each of two
remaining expressions are also independent of time (within our accuracy $O(1/a)$, bearing in mind that the rhs of \rf{3.8} has been multiplied by $a^3$), or they
must cancel each other at any arbitrary moment of time. Can this be the case? The second assumption that the second and third terms on the rhs of Eq.~\rf{3.8} can
cancel each other at any arbitrary moment of time does not work because of the presence of the exponential (with respect to the scale factor $a$) function
$\exp({3{w}_{1}\beta})$ which is linearly independent from the power functions. It can be easily seen that the first assumption does not work as well in the case
${w}_1\neq 0$. For example, we can eliminate the last term in Eq.~\rf{3.8} by putting $\delta\varepsilon_{\mathrm{rad2}}\equiv 0$. However, the second term on the
rhs of this equation still depends on time. The simple analysis shows that, in contrast to Eq. \rf{2.23}, we cannot select the parameters ${w}_{0,1}$ of the model
in such a way that this term becomes independent of time. Already here, we can see a potential problem of the model. However, this term can still be of the order
$o(1/a)$ and then we can drop it within the accuracy of our approach. The problem is that we cannot expand the exponential prefactor into series with respect to
the scale factor which can be arbitrary large. Therefore, we need to estimate the second term taking into account the exponential function. Obviously, this term
is exponentially small for ${w}_1<0$ and $ \beta \gg 1${\footnote{For positive values of ${w}_1>0$, the second term is exponentially divergent when $a\to \infty$.
Hence, there is no need to consider such values since our model must be valid for arbitrary large values of the scale factor $a$.}}. However, we cannot drop the
second term simply on the basis of this argument. To show it, we consider the following ratio:
\ba{3.11} &-&\frac{A \kappa \varphi}{2}a^{-3\left(1+{w}_{0}+{w}_{1}\right)+2} e^{3{w}_{1}\beta}\frac{1}{\frac{\kappa
c^2a^3}{2}\delta\varepsilon_{\mathrm{rad1}}}
\nn \\
&=& -\frac{A \kappa \varphi}{2}a^{-3\left(1+{w}_{0}+{w}_{1}\right)+2} e^{3{w}_{1}\beta}\frac{1}{-\frac{\kappa
c^2a^3}{2}\frac{3\overline{\rho_{\mathrm{c}}}\varphi}{a^4}}= \frac{\overline\varepsilon_{\mathrm{CPL}}}{3\overline{\rho}_{\mathrm{ c}}\, c^2/a^3}\, . \ea
Now, we assume that CPL matter is responsible for the late time acceleration of the Universe. According to the recent observations, it contributes approximately
70\,\% to the total energy density of the Universe, while CDM + baryons give approximately the remaining 30\,\%. Hence, at the present moment the ratio \rf{3.11}
is approximately equal to 0.78. If we consider for ${w}_0$ and ${w}_1$ the values $-0.93$ and $-0.2$ \cite{BOSS14}, respectively, then at the present moment (i.e.
for $\beta \sim 1$) we get that the second term amounts 3\% from the radiation which we take into account in our analysis. This amount is not negligibly small and
we cannot neglect such a time dependent term. Therefore, the considered CPL barotropic fluid with ${w}_1\neq 0$ contradicts the theory of scalar perturbations in
the late Universe.

\section{Dark energy with an EoS parameter linear in $z$}

Now, we consider a model where the  parameter of the EoS for DE is a linear function of the redshift $z$ \cite{Cooray}:
\be{4.1}
\overline p_{z}={w}(z)\overline\varepsilon_{z}\, , \quad
w(z)=w_{0}-w_{1}z=w_{0}+w_{1}\left(1-\frac{a_{0}}{a}\right)\, .
\ee

For this perfect fluid we get
\be{4.2}
\overline\varepsilon_{z}=A a^{-3(1+w_{0}+w_{1})}\, {e}^{-3w_{1}a_0/a}\, .
\ee
By performing an analysis similar to the one carried on the previous section, we conclude that this model also contradicts the theory of scalar perturbations.
Obviously, it happens because of the exponential factor in Eq.~\rf{4.2} which leads to an equation for the comoving gravitational potential similar to \rf{3.8}
where the last two terms on the rhs do not cancel each other.


\section{Dark energy with an EoS parameter linear in $t$}

In the two previous parametrizations of the EoS ${w}$ is a divergent function either for $a \to +\infty$ or for $a \to 0$. In the recent work
\cite{OzAl} a different parametrization for ${w}$ which is based on a linear dependence on time $t$ was proposed:
\be{5.1}
w(t)=w_{0}+w_{1}\left(1-\frac{t}{t_{0}}\right)\, ,
\ee
leading to the following expression in terms of the scale factor $a$:
\ba{5.2} w_{t}(a)&=&w_0+w_1\left[1-\frac{2(1+w_0+w_1)}{w_1+(2+2w_0+w_1) \beta^{-3(1+w_0+w_1)/2}}\right]\nn\\
&=&\frac{-w_{1}\left(w_{0}+w_{1}+2\right)\beta^{3\bar\omega/2}+(2w_{0}+w_{1}+2)(w_{0}+ w_{1})}{w_{1}\beta^{3\bar\omega/2}+2w_{0}+w_{1}+2}\, . \ea
It was shown in \cite{OzAl} that if DE is parametrized  by the EoS \rf{5.2} then: (i) data from Planck, BAO and SN constrain much better the free parameters of
this model (i.e. $w_0$ and $w_1$) as compared with the CPL parametrization, and (ii) if $(w_0,\, w_1) = (-1,\, 1/3)$ in Eq. \rf{5.2}, i.e. the DE EoS parameter
varies between $-2/3$ in the early Universe and $-1$ at present, it fits the data slightly better than a pure $\Lambda$ term. The relation \rf{5.2} was obtained
under the assumption $\bar{w} \neq 0$, where $\bar{w} \equiv w_{0}+w_{1}+1$. More precisely, $0<{w}_1<2\bar{w}$. In general, we can analytically continue this
expression to negative values of $\bar{w}$, keeping the condition $\bar{w}\neq 0$. As we will see below, physical quantities (e.g., the energy density) have the
same asymptotic behavior at big values of the scale factor $a$ for both positive and negative $\bar{w}$. The EoS parameter \rf{5.2} goes to a constant value in
the limit $a\to +\infty$: $w_{t}(a) \to -(\bar{w} +1), \, \bar{w} >0$ and $w_{t}(a) \to -(|\bar{w}|+1)\, ,\, \bar{w} <0$. In the case ${w}_1=0$ we naturally have
$w_{t}(a)={w}_0 = \const$ which was considered at the end of Sec. 2. We note that the direct substitution of $\bar{w}=0$ into Eq.~\rf{5.2} results in an
uncertainty of the type $0/0$. One can check that $w_{t}(a)\rightarrow \frac{-3(w_0+1)\ln(\beta)-4w_0}{3(w_0+1)\ln(\beta)-4}$ as $w_{0}\rightarrow-1-w_{1}$. This
is not a rational fraction and we again will have a problem (except for the trivial case ${w}_0=-1$) similar to the one described in two previous sections.
Therefore, we disregard this case.

The conservation equation for the considered fluid reads
\be{5.3}
 d (\bar{\varepsilon}_{t} \beta^3)+{w}_{t}\bar{\varepsilon}_{t} d (\beta^3)=0\, .
\ee
Now, with the EoS parameter \rf{5.2}, we get the energy density
\be{5.4} \bar{\varepsilon}_{t}=\bar{\varepsilon}_{t_0} \left(1+\frac{w_1\beta^{3\bar{w}/2}-w_1}{2\bar{w}}\right)^4 \beta^{-3\bar{w}}\, , \ee
where $\bar{\varepsilon}_{t_0}$ is an integration constant. It can be easily seen that for both positive and negative $\bar{w}$ the energy density
behaves asymptotically as $\bar{\varepsilon}_{t}\sim \beta^{3|\bar{w}|}\to +\infty$ for $\beta \to +\infty$. In the case ${w}_1=0$ this formula is
reduced to the familiar expression: $\bar{\varepsilon}_{t} = \bar{\varepsilon}_{t_0}\beta^{-3(1+w_0)}$.

Let us turn now to the scalar perturbations for this model. Taking into account Eq.~\rf{5.4}, we get
\be{5.5} \bar{p}_{t}+ \bar{\varepsilon}_{t}=\bar{\varepsilon}_{t}(1+w_{t})=\bar{\varepsilon}_{t_0}\,\bar{w} \beta^{-3\bar{w}}
\left(1+\frac{w_1\beta^{3\bar{w}/2}-w_1}{2\bar{w}}\right)^3 \left(1-\frac{w_1\beta^{3\bar{w}/2}+w_1}{2\bar{w}}\right)\, , \ee
and
\ba{5.6}
&{}&\delta p_{t}=\frac{\partial(w_{t}\bar{\varepsilon}_{t})}{\partial\bar{\varepsilon}_{t}}\delta{\varepsilon}_{t}=\left(w_{t}+
\frac{\partial w_{t}}{\partial a}\frac{\partial a}{\partial \bar{\varepsilon}_{t}}\,\bar{\varepsilon}_{t} \right) \delta{\varepsilon}_{t}\nn\\
&=&\left[\frac{-w_{1}\left(\bar{w}+1\right)\beta^{3\bar{w}/2}+(2\bar{w}-w_{1})(\bar{w}-1)}{w_{1}\beta^{3\bar{w}/2}+2\bar{w}-w_1}+ \frac{w_1
(2\bar{w}-w_{1})\bar{w}}{-w_1^2\beta^{3\bar{w}/2}+(2\bar{w}-w_{1})^2 \beta^{-3\bar{w}/2}}\right] \delta{\varepsilon}_{t}\, . \ea
With the help of Eq.~\rf{2.18} we obtain the expression for $\delta\varepsilon_{t}$:
\ba{5.7} \delta{\varepsilon}_{t}=\frac{-\frac{\varphi}{c^2 a_0}\bar{\varepsilon}_{t_0}\,\bar{w} \beta^{-3\bar{w}-1}
\left(1+\frac{w_1\beta^{3\bar{w}/2}-w_1}{2\bar{w}}\right)^3 \left(1-\frac{w_1\beta^{3\bar{w}/2}+w_1}{2\bar{w}}\right)-
\frac{1}{3}\delta\varepsilon _{\rm rad2}}{\frac{-w_{1}\left(\bar{w}+1\right)\beta^{3\bar{w}/2}+
(2\bar{w}-w_{1})(\bar{w}-1)}{w_{1}\beta^{3\bar{w}/2}+2\bar{w}-w_1}+\frac{w_1 (2\bar{w}-w_{1})\bar{w}}{-w_1^2\beta^{3\bar{w}/2}+
(2\bar{w}-w_{1})^2 \beta^{-3\bar{w}/2}}}\, .
\ea
Then, Eq.~\rf{2.17} takes the form
\ba{5.8}
&{}&\Delta\varphi+3\mathcal{K}\varphi=\frac{\kappa c^4}{2}\delta \rho_{\rm c} \nn\\
&-&\frac{\bar{\varepsilon}_{t_0}\kappa\varphi a_{0}^2}{2} \left[   \frac{\bar {w}\beta^{-3\bar{w}+2}
\left(1+\frac{w_1\beta^{3\bar{w}/2}-w_1}{2\bar{w}}\right)^3
\left(1-\frac{w_1\beta^{3\bar{w}/2}+w_1}{2\bar{w}}\right)}{\frac{-w_{1}\left(\bar{w}+1\right)\beta^{3\bar{w}/2}+(2\bar{w}-w_{1})(\bar{w}-1)}{w_{1}
\beta^{3\bar{w}/2}+2\bar{w}-w_1}+\frac{w_1 (2\bar{w}-w_{1})\bar{w}}{-w_1^2\beta^{3\bar{w}/2}+(2\bar{w}-w_{1})^2 \beta^{-3\bar{w}/2}}}
\right] \nn\\
&+&\frac{\kappa c^2 a_{0}^3 \beta^3}{2} \delta\varepsilon_{\rm
rad2}\left[1-\frac{1}{\frac{-3w_{1}\left(\bar{w}+1\right)\beta^{3\bar{w}/2}+3(2\bar{w}-w_{1})(\bar{w}-1)}{w_{1}\beta^{3\bar{w}/2}+
2\bar{w}-w_1}+\frac{3w_1 (2\bar{w}-w_{1})\bar{w}}{-w_1^2\beta^{3\bar{w}/2}+(2\bar{w}-w_{1})^2 \beta^{-3\bar{w}/2}}} \right]\, .
\ea
It can be easily seen that this equation reduces to Eq.~\rf{2.23} for the particular value ${w}_1=0$ and, consequently, ${w}={w}_0$. Let us now consider the
general case ${w}_1\neq 0$. According to the accuracy of our approach, we should take into account all terms up to the order $O(1/a)$ inclusive (bearing in mind
that the rhs of \rf{5.8} has been multiplied by $a^3$). It should take place for all times in the late Universe including the far future $\beta\gg 1$. In this
limit it is very convenient to expand the expressions under consideration in powers of the scale factor. We see that, because the first line in Eq.~\rf{5.8} is
independent of time (or of the scale factor), all terms of the order $O(1/a^n),\, n\leq 1$ in the next lines must be absent or they must cancel each other. The
term in the second line behaves asymptotically as $\sim \beta^{3|\bar{w}|+2},\,$ when $ \beta\gg 1$, i.e. its asymptotic behavior does not depend on the sign of
$\bar{w}$. The last term behaves asymptotically as $\sim \beta^3 \delta\varepsilon_{\rm{rad2}}, \,$ if $\beta\gg 1$, again for both positive and negative
$\bar{w}$. Therefore, this term has the asymptotic form $\sim 1/\beta$ in the presence of $\delta\varepsilon_{\rm rad2}\sim 1/a^4$ or it equals identically to
zero for $\delta\varepsilon_{\rm rad2}\equiv 0$. Hence, first, in the presence of $\delta\varepsilon_{\rm rad2}$ the second line and the last term cannot cancel
each other, and, second, in the absence of such radiation the time dependent second line cannot be dropped from the equation. This result is in contradiction. We
conclude that the problem has arisen because of the power-law growth of the energy density in the expanding Universe: $\bar{\varepsilon}_{t}\sim
\beta^{3|\bar{w}|}$ when $\beta\gg 1$. Therefore, the model considered in this section contradicts as well the theory of scalar perturbations in the late
Universe.


\section{Conclusions}

In our paper we have studied the late Universe filled with dust-like matter and a barotropic perfect fluid which can play the role of dark energy. We have
considered the radiation content of the Universe as well. More precisely, we have concentrated on scales much smaller than the cell of uniformity size
which is approximately 190 Mpc \cite{EZcosm2}. At such distances our Universe is highly inhomogeneous and the averaged Friedmann approach does not work
here because we need to take into account inhomogeneities in the form of galaxies, groups and clusters of galaxies. It is also a natural assumption that
radiation and the barotropic perfect fluid fluctuate around the average values. Therefore, these fluctuations as well as the dust-like matter
inhomogeneities perturb the FLRW metric. To analyze these perturbations inside the cell of uniformity, we need to use the mechanical approach
\cite{EZcosm1,EKZ2,EZcosm2}. An important feature of this approach is that it provides an optimal tool to study the compatibility of different
cosmological models with observations (see, e.g., \cite{BUZ1,Laslo2}).

In the present paper we have firstly derived the equations for scalar perturbations within the mechanical approach in the case of a barotropic fluid with an
arbitrary EoS $p_{\mathrm{X}}= p_{\mathrm{X}}( \varepsilon_{\mathrm{X}})$ (on top of the dust-like matter and radiation content of the Universe). Our master
equations are Eqs.~\rf{2.17} and \rf{2.18}. First, we have shown how these equations work in the simplest case of a constant EoS parameter $w$ for dark energy.
Our analysis demonstrates that the only two values of $w$, namely $w=-1$ and $w=-1/3$, are acceptable. Therefore, in the case of constant $w$, the cosmological
constant (i.e. $w=-1$) is the only viable dark energy candidate (since a perfect fluid with $w=-1/3$ does not accelerate the Universe).

Then we have applied these equations to variable EoS parameters. Since it is impossible to handle this problem in full generality, we consider three linear
parametrizations of $w$:
CPL (i.e., the parameter of the EoS is linear in the
scale factor $a$), linear in the redshift $z$ and linear in time $t$. Our analysis has shown that all these three models contradict the theory of scalar
perturbations in the late Universe.  This occurs despite the fact that these models can be in sufficient agreement with the observed data obtained from
Planck, BAO and SN \cite{Planck,OzAl,Cooray}.
This means that these EoS are not the fundamental ones and they can be used only for a limited period of time, starting, e.g., from the time of last
scattering and up to the present moment. During this period they approximate/mimic more or less successfully the unknown, ''real'' EoS for DE. In the
future Universe they should be replaced by other expressions. Our investigations show that the most simple and usual constituents of the Universe such as
the cosmological constant, pressureless dark and baryonic matter and radiation are in full agreement with the theory of scalar perturbations.
Interestingly, a perfect fluid with a constant EoS parameter ${w} =-1/3$ fits this scheme also very well \cite{BUZ1}.

We highlight that our conclusion on the incompatibility of perfect fluids which have variable EoS parameters with the theory of scalar perturbations in the late
Universe applies only to the considered cases of linear parametrizations of $w$. In our study, we assumed that these parametrizations take place up to very big
values of the scale factor $a$. This is a rather strong restriction. Therefore, the cases of variable $w$ different from the linear parametrizations require a
separate investigation. For example, it was shown in \cite{Laslo2} that exotic quark-gluon nuggets suitable to describe the dark sides of the Universe
\cite{nugget1,ruth}, if they exist, can also be in agreement with the theory of scalar perturbations while being characterized by variable $w$.


\section*{Acknowledgements}

\"{O}.A. acknowledges the support by T\"{U}B{\.I}TAK Research Fellowship for Post-Doctoral Researchers (2218) and also the support from Ko\c{c} University. The
work of M.B.L. is supported by the Portuguese Agency ``Funda\c{c}\~{a}o para a Ci\^{e}ncia e Tecnologia" through an Investigador FCT Research contract, with
reference IF/01442/2013/CP1196/CT0001. She also wishes to acknowledge the support from the Portuguese Grants PTDC/FIS/111032/2009 and PEst-OE/MAT/UI0212/2014  and
the partial support from the Basque government Grant No. IT592-13 (Spain). The work of M.E. is supported by NSF CREST award HRD-1345219 and NASA grant NNX09AV07A.
A.Zh. acknowledges the hospitality of CENTRA/IST and UBI during the completion of a part of this work.


\section*{Appendix: Adiabatic perturbations of perfect fluids with constant EoS parameters}
\renewcommand{\theequation}{A\arabic{equation}}

In this appendix we obtain the fluctuations of the energy density for perfect fluids with a constant EoS parameter as a function of the scale factor $a$.
For this purpose we consider the conservation equation (see, e.g., Eq. (2.74) in~\cite{Rubakov})
\be{a1}
\delta\varepsilon_{\mathrm{X}}' +3\frac{a'}{a}\left(\delta\varepsilon_{\mathrm{X}}+ \delta p_{\mathrm{X}}\right) + \left(\bar
\varepsilon_{\mathrm{X}}+\bar p_{\mathrm{X}}\right)(\Delta v-3\Phi')=0\, ,
\ee
where $\Delta$ is the Laplace operator with respect to the comoving coordinates and $v$ is the comoving peculiar velocity potential. We remark that we consider
the fluctuations of perfect fluids arising due to inhomogeneities of dust-like matter (e.g., galaxies). These fluctuations also form their own inhomogeneities.
Then the velocity potential $v$ characterizes dynamics of displacement of such inhomogeneities. For example, in the case of radiation, $v$ corresponds not to the
speed of light but to the speed of displacement of photon concentrations. In the late Universe these displacement speeds are nonrelativistic and give much less
contributions than $\Phi$ \cite{EZcosm1,EZcosm2}. Therefore, we drop the term containing $v$ in Eq. \rf{a1}. The gravitational potential is given by the formula
\rf{2.7}: $\Phi=\varphi/(c^2a)$. We consider perfect fluids with a constant EoS parameter ${w}$: $\bar p_{\mathrm{X}}={w} \bar \varepsilon_{\mathrm{X}}$. Hence,
the conservation equation gives $\bar \varepsilon_{\mathrm{X}}=\bar A/a^{3(1+{w})}$. For such fluids their adiabatic fluctuations also satisfy the same EoS:
$\delta p_{\mathrm{X}} = {w} \delta\varepsilon_{\mathrm{X}}$. Then the general solution of Eq.~\rf{a1} may be written in the form
\be{a2} \delta\varepsilon_{\mathrm{X}}=\frac{A}{a^{3(1+{w})}}+\frac{3\varphi(1+{w})\bar A}{c^2a^{3(1+{w})+1}}\, ,\ee
where $A$ is a constant of integration. This expression shows that in the case of the negative parameter ${w}<0$ both terms in the rhs satisfy the adopted
accuracy because $3(1+{w})+1<4$. However, the Einstein equations (in our case, in the form of Eqs. \rf{2.17} and \rf{2.18}) may impose additional
restrictions on the form of $\delta\varepsilon_{\mathrm{X}}$. For example, the agreement with Eq. \rf{2.22} where $\delta\varepsilon_{\mathrm{rad2}}=0$
results in additional conditions: $A=0,\, -(1+{w})/{w} = 3(1+{w})$. The latter condition holds true only for ${w}=-1$ and ${w}=-1/3$.
Therefore,
\be{a8} \delta \varepsilon_{-1/3}=\frac{2\varphi}{c^2}\bar A \frac{1}{a^3}\, . \ee
The conclusion that the value ${w}=-1/3$ is a particular one was also made from different reasoning in papers by Fulvio Melia (see, e.g., \cite{Melia}
and the references therein). It can be also easily seen that in the case ${w}=0$ the formula \rf{a2} exactly reproduces Eq. \rf{2.8} if we put
$A=\delta\rho_{\mathrm{c}}\, c^2$, $\bar A=\bar\rho_c c^2$. In the case of radiation ${w}=1/3$ the second term in \rf{a2} is of the order $O(1/a^5)$
and should be dropped. Then, we get
\be{a5} \delta \varepsilon_{\mathrm{rad}} \sim 1/a^4\, . \ee

It is worth mentioning that in the general case of interacting fluids Eq. \rf{a1} must contain their total energy densities and pressures. This case is
beyond the scope of our paper where we concentrated exclusively on the standard perfect fluids without energy exchange between them.

There is also one more conservation equation (see, e.g., (2.75) in \cite{Rubakov}).
In our approach (that is when neglecting the peculiar velocities) this equation reads
\be{a9} \sum_{i}\left[\delta p_{\mathrm{X}_i} + \left(\bar \varepsilon_{\mathrm{X}_i}+\bar p_{\mathrm{X}_i}\right)\Phi\right]=0\, . \ee
Let us consider a model with four perfect fluids: dark energy represented by the cosmological constant, CDM, radiation and the ${w} =-1/3$ fluid. Then,
\rf{a9} takes the following form:
\ba{a10} &{}& \delta p_{\Lambda} + \left(\bar \varepsilon_{\Lambda}+\bar p_{\Lambda}\right)\Phi +
\delta p_{\mathrm{CDM}} + \left(\bar \varepsilon_{\mathrm{CDM}}+\bar p_{\mathrm{CDM}}\right)\Phi \nn \\
&+& \delta p_{\mathrm{rad}} + \left(\bar \varepsilon_{\mathrm{rad}}+\bar p_{\mathrm{rad}}\right)\Phi + \delta p_{-1/3} + \left(\bar
\varepsilon_{-1/3}+\bar p_{-1/3}\right)\Phi = 0\, . \ea
Because $\delta p_{\Lambda} =0$ and $\bar \varepsilon_{\Lambda} = -\bar p_{\Lambda}$, the contribution from the $\Lambda$-term disappears. Now, taking
into account that $\delta p_{\mathrm{CDM}} = \bar p_{\mathrm{CDM}}=0\, ,\; \delta p_{\mathrm{rad}} = (1/3)\delta \varepsilon_{\mathrm{rad}}$ and
$\left(\bar \varepsilon_{\mathrm{rad}}+\bar p_{\mathrm{rad}}\right)\Phi \sim O(1/a^5)$, we get
\be{a11} \delta \varepsilon_{\mathrm{rad}} = -3 \bar \varepsilon_{\mathrm{CDM}} \Phi = -\frac{3\overline{\rho}_{\mathrm{c}}\,\varphi}{a^4}\, , \ee
where $\bar \varepsilon_{\mathrm{CDM}}= \bar \rho_{\mathrm{c}}c^2/a^3$. This expression exactly coincides with the formula \rf{2.10} in Sec. 2. Finally,
keeping in mind that $\bar p_{-1/3} = -(1/3)\bar \varepsilon_{-1/3}$ and $ \delta p_{-1/3} = -(1/3)\delta \varepsilon_{-1/3}$, we arrive at Eq. \rf{a8}.
Therefore, we have shown the self-consistency of our approach.


\end{document}